\documentclass[12pt]{article}
\usepackage{latexsym}
\usepackage{amsfonts}

\catcode`\@=11

\def\section{\@startsection{section}{1}{\z@}{3.5ex plus 1ex minus
 .2ex}{2.3ex plus .2ex}{\bf}}

\def\thesubsection{\arabic{section}.\arabic{subsection}}
\renewcommand{\subsection}[1]{\addtocounter{subsection}{1}
\vspace{2.5mm}\par\noindent {\it \thesubsection . #1}\par
 \vspace{0.5mm} }
\catcode`\@=12
\newfont{\mbm}{msbm10 scaled\magstep1}

\DeclareFontFamily{U}{rsf}{}
\DeclareFontShape{U}{rsf}{m}{n}{
  <5> <6> rsfs5 <7> <8> <9> rsfs7 <10-> rsfs10}{}
\DeclareMathAlphabet\Scr{U}{rsf}{m}{n}

\mathchardef\varGamma="0100
\mathchardef\varDelta="0101
\mathchardef\varTheta="0102
\mathchardef\varLambda="0103
\mathchardef\varXi="0104
\mathchardef\varPi="0105
\mathchardef\varSigma="0106
\mathchardef\varUpsilon="0107
\mathchardef\varPhi="0108
\mathchardef\varPsi="0109
\mathchardef\varOmega="010A

\def\drawbox#1#2{\hrule height#2pt\hbox{\vrule width#2pt height#1pt
 \kern#1pt\vrule width#2pt}\hrule height#2pt}

\def\Asym#1#2{\vcenter{\vbox{\drawbox{#1}{#2}\kern-#2pt\drawbox{#1}{#2}}}}

\begin{document}
\begin{titlepage}
\rightline{{CERN-TH/2003--248}}
\rightline{{SPIN-03/31}}
\rightline{{ITP-03/50}}
\rightline{{UCLA/03/TEP/28}}
\rightline{{hep-th/0310136}}
\vskip 2cm
\centerline{{\large\bf Unusual gauged supergravities}}
\vskip .2cm
\centerline{{\large\bf
from type IIA and type IIB orientifolds}}
\vskip 1cm
\centerline{Carlo Angelantonj${}^1$, Sergio Ferrara${}^{1,2,3}$ and
Mario Trigiante${}^4$}
\vskip 0.5cm
\centerline{\it ${}^1$ Theory Division --- CERN, CH-1211 Geneva 23}
\vskip 0.2cm
\centerline{\it ${}^2$ INFN, Laboratori Nazionali di Frascati, Italy}
\vskip 0.2cm
\centerline{\it ${}^3$ Dept. of Physics and Astronomy, University of
California, Los Angeles}
\vskip 0.2cm
\centerline{\it ${}^4$ Spinoza Institute, Leuvenlaan 4, NL-3508 Utrecht}
\vskip  1.0cm
\begin{abstract}
We analyse different ${\Scr N} =4$ supergravities coupled to 
six vector multiplets corresponding to low-energy descriptions of the 
bulk sector of $T_6/\mathbb{Z}_2$ orientifolds with $p$-brane in IIB 
($p$ odd) and in IIA ($p$ even) superstrings.
When fluxes are turned on, a gauging emerges corresponding to some 
non-semisimple Lie algebra related to nilpotent subalgebras $N_p \subset 
{\rm so} (6,6)$, with dimension $h_{N_p} = 15 + (p-3)(9-p)$. 
The non-metric axions have Stueckelberg couplings that induce a 
spontaneous breaking of gauge symmetries. In four cases the gauge algebra 
is non-abelian with a non-commutative structure of the compactification 
torus, due to fluxes of NS-NS and R-R forms.
\end{abstract}
\end{titlepage}

\section{Introduction}

Effective four-dimensional supergravity theories obtained by superstring
compactifications on certain six-dimensional manifolds are not only distinct
by the number of supersymmetries preserved by the background, but also by
the duality symmetries which act linearly on the vector fields.
Although in general, theories with the same amount of supersymmetries are 
related by a (non-local) symplectic change of the duality basis acting
on the electric and magnetic field strengths \cite{Gaillard:1981rj}, 
after some isometries are 
gauged, that in theories with ${\Scr N} >1$ also amounts to the generation
of a scalar potential, such change of basis is no longer allowed, and different
gaugings describe genuinely different vacua 
\cite{Andrianopoli:2002aq,Dabholkar:2002sy,deWit:2002vt}.

The simplest manifestation of this phenomenon is perhaps given by two different
gaugings of ${\Scr N}=8$ four-dimensional supergravity
\cite{Andrianopoli:2002aq}: the SO(8) gauging \cite{deWit:1981eq}, 
corresponding to M-theory on ${\rm AdS}_4 \times 
S_7$, and the ${\Scr N}=8$ spontaneously broken supergravity 
dimensionally reduced {\it \`a la} Scherk--Schwarz \cite{SS}
on ${\Scr M}_4
\times T_7$. In the former case the gauge algebra is a subalgebra of
${\rm sl} (8,\mathbb{R}) 
\subset {\rm e}_{7,7}$, while in the latter example the
``flat algebra'' is a subalgebra of $({\rm e}_6 + {\rm so}(1,1))
+ T_{27} \subset {\rm e}_{7,7}$ \cite{Andrianopoli:2002mf}.

Similar manifestations also appear in ${\Scr N}=4$ supergravities describing
$T_6 /\mathbb{Z}_2$ orientifolds , where the $\mathbb{Z}_2$ 
projection is a combination
of the world-sheet parity $\varOmega$ and geometric inversions of 
$9-p$ directions of the compactification six-torus \cite{cargese,orient,horava,
pol, revs}. 
Indeed, in the two
extremal cases of IIB orientifolds with $p=3$ and $p=9$ one is led to 
completely different low-energy supergravities. In the former case the
fifteen Peccei-Quinn symmetries of the $C_{MNPQ}$ R-R scalars do not rotate
the twelve vectors $B_{\mu i}$ and $C_{\mu i}$, and thus can be gauged
\cite{fp,kst,D'Auria:2003jk,Berg:2003ri}
yielding a twelve-dimensional abelian gauge algebra. On the other hand, the
$p=9$ case corresponds to the $T_6$ reduction of the ${\Scr N}=1$
ten-dimensional type I superstring. The fifteen Peccei-Quinn symmetries of the
$C_{MN}$ R-R scalars now rotate the twelve vectors ${\Scr G}_\mu^i$ and 
$C_{\mu i}$
\begin{equation}
\delta C_{\mu i} = \xi_{ij} {\Scr G}_\mu ^j \,,
\end{equation}
and no gauging is thus possible. The other orientifolds with $3<p<9$
appear as intermediate cases of these two, with the twelve vectors
originating in part by the metric $G_{MN}$, in part by the NS-NS $B$-field, 
and in part by the R-R $C$-forms \cite{Angelantonj:2003rq}.

When fluxes are turned on \cite{ps}--
\cite{Tripathy:2003qw} (see \cite{Frey:2003tf} for a comprehensive review), 
a very rich structure emerges depending
on $p$. In particular, for $4<p<9$, the $p-3$ graviphotons 
${\Scr G}_{\mu}^i$ always gauge ``non-abelian''
isometries when the $H$-flux of the $B$ field strength is
non-vanishing. This is a new manifestation of a non-commutative
structure of the compactification torus in the presence of a non-trivial NS-NS
background. For each case, there is a non-injective 
homomorphism $\iota $ between the gauge group ${\Scr G}_g$, under which the 
gauge fields transform in the adjoint representation, and its realisation 
${\Scr G}^\prime_g$ in terms of isometries of the scalar manifold, which is fixed by the scalar--vector minimal couplings:
\begin{eqnarray}
{\Scr G}_g &\stackrel{\iota}{\longrightarrow }& 
{\Scr G}^\prime_g\subset {\rm Isom}({\Scr M}_{scal}) \,,
\nonumber\\
{\Scr G}^\prime_g &\equiv & {\Scr G}_g/ {\rm Ker}(\iota) \qquad {\rm with} 
\qquad {\rm Ker}(\iota)\,\neq\, \emptyset \,.
\label{iota}
\end{eqnarray}
Elements in ${\rm Ker}(\iota)$ are central charges in
the gauge algebra $\mathbb{G}_g$ of ${\Scr G}_g$ whose action is trivial on 
the scalar fields, and amounts to a pure \emph{gauge} transformation on
the vector fields. In some cases, the closure of ${\Scr G}'_g$ requires
additional conditions on the fluxes.

The structure of the gauge algebras for the IIB orientifolds with 
$p=7$ and $p=5$, originally outlined in \cite{Angelantonj:2003rq}, 
where also the salient features 
of the underlying (ungauged) supergravities were exposed, is here summarised 
in section 2. Section 3 contains new results on the gauge algebras emerging 
from IIA orientifolds ($p$ even). Finally, in section 4 our conclusions are 
drawn.

\section{The gauge algebra of IIB orientifolds with fluxes}

We recall here the gauge algebras of IIB orientifolds with $p=7$ and $p=5$,
first exploited in \cite{Angelantonj:2003rq}. 
To fix the notation, it is convenient to split the six-torus as
\begin{equation}
T_6 = T_{p-3}\times T_{9-p} \,, \label{torusdec}
\end{equation}
with indices $i,j = 1,\ldots ,p-3$ labeling coordinates along the $T_{p-3}$
sub-torus, and indices $a,b = 1, \ldots , 9-p$ labeling the coordinates in
$T_{9-p}$. The $\mathbb{Z}_2$ symmetry we are implementing is a combination
of world-sheet parity $\varOmega$ and inversions $I_{9-p}$
of the $9-p$ coordinates $y^a$ of $T_{9-p}$. As a result, only the
subgroup ${\rm GL} (p-3) \times {\rm GL} (9-p)$ of the isometries of the
six-torus is perturbatively realised in the orientifold models we are 
interested in, and thus the decomposition (\ref{torusdec}) turns out to be
the natural one.

\subsection{The $T_4 \times T_2$ model}

In this model the bulk gauge fields and the non--metric axions, invariant under the $\varOmega I_4$ projection, are:
\begin{eqnarray}
&&{\Scr G}_\mu^i,\,B_{\mu a},\,C_{\mu a},\,C_\mu^i = \epsilon^{ijkl}
C_{\mu jkl}\,,\nonumber\\
&&
 C_0,\,B_{ia},\,C_{ia},\,C_{ijab}=C_{ij}\,\epsilon_{ab},\,C_{ijk\ell} \,.
\end{eqnarray}
We shall focus on the effect of the fluxes
\begin{eqnarray}
F_{ija},\,H_{ija},\, G_{ijk ab}\,,
\end{eqnarray}
where $F_{ija},\,H_{ija}$ are the R--R and NS--NS three-form fluxes
 while $G_{ijk ab}$ is the flux of the five--form field--strength, whose effect was not considered in our previous analysis \cite{Angelantonj:2003rq}.
For our purposes it is convenient
to collect the $B_{\mu a}$ and $C_{\mu a}$ vectors as well as the $B_{ia}$ and $C_{ia}$ scalars and the fluxes $H_{ija}$ and $F_{ija}$ into SO(2,2) covariant quantities:
$A_\mu^\varLambda$, $\varPhi^\varLambda_i$ and $H^\varLambda_{ij}$ ($\varLambda =1,\ldots , 4$). The $C_\mu^i$ vectors
decouple completely so that the active gauge algebra $\mathbb{G}_g$ of 
${\Scr G}_g$ is eight-dimensional with connection 
\begin{equation}
\varOmega_g = X_i {\Scr G}^i_\mu + X_\varLambda A_\mu^\varLambda
\,,
\end{equation}
and with the following structure constants
\begin{equation}
[ X_i , X_j ] = H^\varLambda_{ij} X_\varLambda\,, \qquad [X_i ,
X_\varLambda ] = [ X_\varLambda , X_\varSigma ]= 0\,. \label{alg71}
\end{equation}
On the other hand, there are $15 + (p-3)(9-p)$ (twenty-three in
this case) scalar axions, whose associated solvable subalgebra 
\cite{Andrianopoli:1996bq,Andrianopoli:1996zg,Cremmer:1997ct,
Cremmer:1998px} of so(6,6) is \cite{Angelantonj:2003rq}
\begin{equation}
[T_0 , T^i_\varLambda ] = {\Scr M}_\varLambda {}^{\varLambda '}
T^i_{\varLambda '} \,, \qquad [T_\varLambda^i , T_{\varLambda '}^j
] = \eta_{\varLambda \varLambda '} T^{ij} \,, \label{alg72}
\end{equation}
with the remaining commutators vanishing. 
The realisation ${\Scr G}'_g$ of the gauge algebra in terms of isometries
of the scalar manifold is achieved through the following identification of its generators:
\begin{eqnarray}
X^\prime_i &=& -H^\varLambda_{ij} T^{j}_{\varLambda}+G_{ijkab}\, T^{jk}\,,
\nonumber \\
X^{\prime \varLambda} &=& {\textstyle{1\over
2}} H^\varLambda_{ij} T^{ij} \,. \label{ident}
\end{eqnarray} 
 Notice that the presence of the five--form flux $G_{ijkab}$ does not affect the structure of the gauge algebra but amounts to an additional term in the covariant derivative of $C_{ij}$:
\begin{equation}
D_\mu C_{ij} = \partial_\mu
C_{ij}-{\textstyle\frac{1}{2}}\,H_{ij\varLambda}\,
A^\varLambda_\mu-{\Scr G}^k_\mu\,
G_{kijab}+{\textstyle\frac{1}{2}}\,{\Scr
G}^k_\mu\,H^\varLambda_{k[i}\,\varPhi_{j]\varLambda} \,.
\end{equation}

In general, the identification of the gauge generators with isometries
does not guarantee automatically that the gauge algebra $\mathbb{G}'_g$
be compatible with $\mathbb{G}_g$. Indeed, in the case at hand,
one can show that the expressions (\ref{ident}) for the generators of
$\mathbb{G}'_g$ reproduces the
structure (\ref{alg71}) of $\mathbb{G}_g$ only if the following 
condition on the fluxes is fulfilled:
\begin{equation}
H^{\varLambda}_{ij} H^{ij}_\varLambda =0\,.
\end{equation}
This is consistent with the fact that the theory contains
seven-branes ($p=7$). Interestingly enough, this condition also
allows a lift of the ${\Scr N} =4$ theory to a truncation of a
${\Scr N} =8$ gauge algebra \cite{toappear}.

\subsection{The $T_2 \times T_4$ model}

In this example \cite{Angelantonj:2003rq} 
the twelve vector fields and the non--metric axions which are invariant under the orientifold projection  are:
\begin{eqnarray}
&&{\Scr G}_\mu^i,\,B_{\mu a},\,C_{\mu i},\,C_\mu^a = \epsilon^{abcd} C_{\mu bcd},\,\nonumber\\
&&C_{ab},\,B_{ia},\,C_i^a=\epsilon^{abcd}\, C_{ibcd},\,C_{\mu\nu},\,C_{ij}\,.
\end{eqnarray}
 Also in this case the $C_{\mu i}$
decouple, so that the active gauge algebra is ten-dimensional, with connection
\begin{equation}
\varOmega_g = {\Scr G}_\mu ^i X_i + B_{a\mu} X^a + C_\mu^a X_a\,.
\end{equation}
We shall consider only the effect of the NS-NS and R-R three-form fluxes $H_{ija} = \epsilon_{ij} H_a$ and $F_{iab}$. They appear as structure constants in the gauge
algebra
\begin{equation}
[X_i , X_j] = \epsilon_{ij} H_a X^a \,, \qquad [X_i , X^a ] =
F_{i}{}^{ab} X_b \,, \label{alg24}
\end{equation}
with the remaining commutators vanishing\footnote{Indices are lowered and 
raised with the $\epsilon_{ij}$ and $\epsilon_{abcd}$ tensors}.

Turning to the scalar sector,
the generators $T$, $T^{ia}$, $T^i_a$ and $T^{ab}$ of the 
twenty-three dimensional solvable algebra $N_5$ associated to the 
relevant axionic non--metric scalars obey the commutation relations
\begin{equation}
[T^{ia},T^{bc} ] = \epsilon^{abcd} T^i_d \,, \qquad [T^{ia},
T^j_b] = \epsilon^{ij} \delta^a_b \,.
\end{equation}
One is thus led to the following identifications
\begin{equation}
X'_i = - F_i {}^{ab} T_{ab} + H_a T^{a}_i \,, \qquad X'_a = - H_a
T \,, \qquad X^{\prime a} = F_i {}^{ab} T^i_b \,,
\end{equation}
of the gauge generators with the isometries of the solvable algebra.
However, they 
reproduce now only a contracted version of $\mathbb{G}_g$ as given in
(\ref{alg24}). Indeed, as we have already stated, 
the groups ${\Scr G}_g$ and ${\Scr G}_g^\prime$ are 
related by the non-injective homomorphism
(\ref{iota}), where now ${\rm Ker}(\iota)$ 
is generated by the three central charges $X_a$ orthogonal to $X'_a$.

Moreover, no further constraints are to be imposed on the fluxes, 
that however satisfy $H_3 \wedge F_3 =0$ identically, 
at all consistent with the fact that the model would now include D5 branes.
Also this model can be lifted to a gauged ${\Scr N}=8$ theory \cite{toappear}.

\section{Type IIA orientifolds}

We now turn to the description of gauge algebras of IIA orientifolds 
with fluxes, for the three different cases $p=8,6$ and 4. Their spectra and 
ungauged low-energy supergravities have already been discussed in 
\cite{Angelantonj:2003rq}.

\subsection{The $T_5\times T_1$ model}

Aside from the four-dimensional graviton $g_{\mu\nu}$, and the geometric 
moduli $g_{ij}$ and $g_{99}$ of $T_5 \times T_1$, the massless bosonic
spectrum consists of 
\begin{eqnarray}
\mbox{scalars (axionic):}&&
C_i,\,B_{i9},\,C_{ij9},\,C_{\mu\nu 9} \,,
\nonumber\\
\mbox{vector fields:}&&  {\Scr G}^i_{\mu},\,C_{i9\mu},\,C_\mu
,\,B_{9\mu} \,,
\end{eqnarray}
while only the $H_{ij9}$ and $G_{ijk9}$ fluxes for the NS-NS $B$-field and R-R
three-form potential are allowed by the orientifold projection.

The gauge group ${\Scr G}_g$ is generated by the algebra $\mathbb{G}_g =
\{ X_i,\,X,\,X^{i9},\,X^9\}$, with connection
\begin{equation}
\varOmega^g = {\Scr G}^i_\mu\,X_i+C_\mu\,X+C_{i9\mu}\,X^{i9}+B_{9\mu}\,X^9 \,.
\end{equation}
When fluxes are turned on, they appear as structure constants in the 
commutators
\begin{equation}
\left[X_i,\,X\right] = -H_{ij9}\,X^{j9}\, , 
\qquad 
\left[X_i,\,X_j\right] = H_{ij9}\,X^9+G_{ijk9}\,X^{k9},
\label{51alg}
\end{equation}
from which we deduce that the generators $\{X^9,\,X^{i9}\}$ are central 
charges. The form of the algebra (\ref{51alg}) then suggests that the field
strength of the vector fields present non-abelian couplings
\begin{eqnarray}
{\Scr F}^i_{\mu\nu} &=& \partial_\mu {\Scr G}^i_\nu-\partial_\nu 
{\Scr G}^i_\mu \,,
\nonumber\\
F_{i9\mu\nu} &=& \partial_\mu C_{i9\nu}-\partial_\nu
C_{i9\mu} + {\Scr G}_\mu^k \,C_\nu\, H_{ki9} - {\Scr G}_\nu^k \,C_\mu\,
H_{ki9} - {\Scr G}_\mu^k \, {\Scr G}^\ell_\nu\, G_{k\ell i9} \,,
\nonumber\\
F_{\mu\nu} &=& \partial_\mu C_{\nu}-\partial_\nu C_{\mu} \,,
\nonumber\\
{\Scr H}_{9\mu\nu} &=& \partial_\mu B_{9\nu}-\partial_\nu B_{9\mu}
- {\Scr G}_\mu^k \, {\Scr G}^\ell_\nu\, H_{k\ell 9} \,,
\end{eqnarray}
as is confirmed by a supergravity inspection.

Turning to the scalar sector, we have shown in \cite{Angelantonj:2003rq} 
that the solvable
algebra parametrised by the (non-metric) axionic scalars is generated by
\begin{equation}
N_8 = \{ B_{i9} T^{\prime \, i} + C_i T^i + C_{ij9} T^{ij} \} \,,
\end{equation}
with the only non-vanishing commutator given by
\begin{equation}
[T^i , T^{\prime j} ] = T^{ij} \,.
\end{equation}

The group ${\Scr G} ' _g$ of gauge transformations on the axionic scalars 
is now generated by the algebra $\mathbb{G} '_g = \{ X^\prime_i\,,\,
X^\prime\}$, since in this case  ${\rm Ker}(\iota)\,=\,\{X^9,\,X^{i9}\}$. 
The realisation of $\mathbb{G}'_g$ in the terms of isometries of the 
scalar manifold suggests the identifications
\begin{eqnarray}
X^\prime_i &=& H_{ij9}\, T^{\prime j}-G_{ijk9}\, T^{jk}\,,
\nonumber\\
X^\prime &=& H_{ij9}\,T^{ij} \,, \label{51gen}
\end{eqnarray}
that reproduce the structure (\ref{51alg}) once we set to zero the 
central charges.

The generators (\ref{51gen}) induce then the following transformations
on the scalars
\begin{eqnarray}
\delta \tilde{C}_{ij9} &=& -\xi\,
H_{ij9}-\xi^k\,G_{ijk9}+\xi^k\,H_{k[i|9}\,C_{j]} \,,
\nonumber\\
\delta B_{i9} &=& \xi^j\,H_{ji9}\,,
\nonumber\\
\delta C_i &=& 0 \,,
\nonumber\\
\delta C_{\mu\nu 9} &=& 0 \,,
\end{eqnarray}
where we have found convenient to define the scalar
$C_{ij9} \to \tilde C _{ij9} = C_{ij9} - C_{[i} B_{j]9}$. 
As a result, the corresponding covariant derivatives read
\begin{eqnarray}
D_\mu \tilde{C}_{ij9} &=& \partial_\mu \tilde{C}_{ij9} + C_\mu\,
H_{ij9} + {\Scr G}^k_\mu\, G_{ijk9} - {\Scr G}^k_\mu\,H_{k[i|9}\,C_{j]}\,,
\nonumber\\
D_\mu B_{i9} &=& \partial_\mu B_{i9} - {\Scr G}^k_\mu\,H_{ki9}\,,
\nonumber\\
D_\mu C_{i}&=&\partial_\mu C_i \,.
\end{eqnarray}

\subsection{The $T_3\times T_3$ model}

The next model we shall describe, is the $T_3 \times T_3 /\mathbb{Z}_2$
orientifold of the IIA superstring. Its massless spectrum comprises, aside from
the four-dimensional metric $g_{\mu\nu}$, 
the vector fields
\begin{equation}
 {\Scr G}_\mu^i \,,\, C_{ij\mu} \,,\, B_{a\mu} \,,\, C_{ab\mu}  \,,
\end{equation}
the dilaton,
the geometric moduli $g_{ab}$ and 
$g_{ij}$ of the six-torus in its $T_3 \times T_3$ decomposition, and the
axionic scalars $\{C^{ab},\,B_{ia},\,C_{iab},\,C_{k\mu\nu }=C_{ij},\,
C_{ijk}\}$. These latter, aside from $C_{ijk}$, parametrise a twenty-four dimensional solvable
subalgebra
\begin{equation}
N_6 = \{B_{ia}\,T^{i a}+C^{ab}\,T_{ab}+C_{i}^a\,
T^{i}_a+C_{ij}\,T^{ij}\}\,,
\end{equation}
whose structure is encoded in the non-vanishing commutators
\begin{eqnarray}
\left[T_{ab},\,T^{ic}\right] &=& T^{i}_{[a}\delta^c_{b]} \,,
\nonumber\\
\left[T^{ia},\,T^{j}_b\right] &=& T^{ij} \delta^a_b \,.
\end{eqnarray}

The active gauge group ${\Scr G}_g$ is generated by the algebra 
$\mathbb{G}_g\,=\,\{X_i,\,X^a,\,X^{ab}\}$ with connection
\begin{equation}
\varOmega^g = {\Scr G}^i_\mu\,X_i+C_{ab\mu}\,X^{ab}+B_{a\mu}\,X^a \,.
\end{equation}
We shall consider the effect of the fluxes
\begin{equation}
F_{ia} \,,\, H_{ija} \,,\, G_{ijab}\,,
\end{equation}
which determine a  non-abelian gauge algebra, with commutators
\begin{eqnarray}
\left[X_i,\,X_j\right] &=&
H_{ija}\,X^{a} + G_{ijab}\,X^{ab}\,,
\nonumber \\
\left[X^a,\,X_i\right] &=& {\textstyle\frac{1}{2}}\,
F_{ib}\,X^{ab} \,. \label{33alg}
\end{eqnarray}
As a result the field strengths of the vector fields read
\begin{eqnarray}
{\Scr H}_{a\mu\nu } &=& \partial_\mu B_{a\nu} - \partial_\nu
B_{a\mu}- {\Scr G}^i_\mu {\Scr G}^j_\nu\,H_{ij a}\,,
\nonumber\\
F_{ij\mu\nu } &=& \partial_\mu C_{ij\nu} - \partial_\nu C_{ij\mu}\,,
\nonumber\\
F_{ab\mu\nu } &=& \partial_\mu C_{ab\nu} - \partial_\nu
C_{ab\mu} - {\Scr G}^i_\mu {\Scr G}^j_\nu\, G_{ijab} - 
{\textstyle\frac{1}{2}}\,{\Scr G}^i_\mu \,
F_{i[a}\, B_{b]\nu} + {\textstyle\frac{1}{2}}\, {\Scr G}^i_\nu \, F_{i[a}\, 
B_{b]\mu}\,,
\nonumber\\
{\Scr F}^i_{\mu\nu} &=& \partial_\mu {\Scr G}^i_{\nu} - \partial_\nu
{\Scr G}^i_{\mu} \,.
\end{eqnarray}

The group ${\Scr G}^\prime_g$ of gauge transformations on the axionic scalars
is generated by the algebra $\mathbb{G}^\prime_g\,=\,\{X^\prime_i,\,
X^{\prime a},\,X^{\prime ab}\}$, and is realised in terms of isometries of 
the scalar manifold by the identifications
\begin{eqnarray}
X_i^\prime &=& - {\textstyle\frac{1}{4}}\,\epsilon^{abc}\,F_{ia}\,
T_{bc}+H_{ija}\, T^{ja}+ {\textstyle\frac{1}{2}}\,G_{ijab}\,T^{j}_c \,,
\nonumber\\
X^{\prime a} &=& {\textstyle\frac{1}{4}}\,\epsilon^{abc}\,G_{bcij}\,T^{ij}+
{\textstyle\frac{1}{4}}\,\epsilon^{abc}\,F_{ib}\,T^i_c \,,
\nonumber\\
X^{\prime ab} &=& {\textstyle\frac{1}{4}}\,\epsilon^{abc}\,H_{ijc}\,T^{ij} \,.
\label{identi}
\end{eqnarray}
An explicit calculation of their commutators, then shows that the algebra
$\mathbb{G}'_g$ reproduces the structure (\ref{33alg}) of $\mathbb{G}_g$
if the following conditions on the fluxes are met
\begin{equation}
V^c = \epsilon^{ijk}\,\epsilon^{abc}\,F_{ia}\, H_{jkb}\,=\,0 \,,
\end{equation}
that also imply the useful relation
\begin{equation}
\epsilon^{abc}\,F_{[i|a}\,
H_{j]kb} = -{\textstyle\frac{1}{2}}\,F_{ka}\,H_{ijb} \,.
\end{equation}

The identifications (\ref{identi}) induce the following gauge transformations
on the axionic scalars
\begin{eqnarray}
\delta C^a_i &=& {\textstyle\frac{1}{4}}\, \epsilon^{abc}\,\xi_b\,
F_{ic} + {\textstyle\frac{1}{2}}\, \epsilon^{abc}\, \xi^j\, G_{jibc}+
{\textstyle\frac{1}{4}}\,\epsilon^{abc}\,\xi^j\, F_{ja}\, B_{ic} \,,
\nonumber\\
\delta C_a &=& -{\textstyle\frac{1}{2}}\,\xi^i\, F_{ia} \,,
\nonumber\\
\delta B_{ia} &=& \xi^j\,H_{jia}\,,
\nonumber\\
\delta C_{ij}&=& {\textstyle\frac{1}{4}}\,\epsilon^{abc}\,\xi_{ab}\,H_{ij
c}+{\textstyle\frac{1}{4}}\,\epsilon^{abc}\,\xi_a\,G_{ij
bc}-\xi^k\,H_{k[i|a}\,C_{j]}^a \,,
\end{eqnarray}
that generate the minimal couplings
\begin{eqnarray}
D_\mu C_{i}^a &=& \partial_\mu C_{i}^a - {\textstyle\frac{1}{4}}\,
\epsilon^{abc}\, B_{b\mu}\, F_{ic} - {\textstyle\frac{1}{2}}\, 
\epsilon^{abc}\, {\Scr G}^j_\mu\, G_{jibc} -
{\textstyle\frac{1}{4}}\,\epsilon^{abc}\, {\Scr G}^j_\mu\, F_{ja}\, B_{ic}\,,
\nonumber\\
D_\mu C_a &=& \partial_\mu  C_a + {\textstyle\frac{1}{2}}\,
{\Scr G}_\mu^i\, F_{ia}\,,
\nonumber\\
D_\mu B_{ia} &=& \partial_\mu B_{ia} - {\Scr G}_\mu^j\,H_{jia}\,,
\nonumber\\
D_\mu C_{ij} &=& \partial_\mu C_{ij}-{\textstyle\frac{1}{4}}\,
\epsilon^{abc}\,C_{ab\mu}\,H_{ijc} -
{\textstyle\frac{1}{4}}\,\epsilon^{abc}\, B_{a\mu}\,G_{ijbc} + 
{\Scr G}^k_\mu\,H_{k[i|a}\,C_{j]}^a \,.
\end{eqnarray}

\subsection{The $T_1\times T_5$ model}

Finally, we consider the $T_1 \times T_5$ orientifold. The relevant bosonic
fields are
\begin{eqnarray}
\mbox{scalars (axionic):}&&
 C_{abc},\,B_{4a},\,C_{a\mu\nu }=C^b,\,C_{4}\,,
\nonumber\\
\mbox{vector fields:}&&
{\Scr G}^4_{\mu},\,C_{\mu},\,C_{4a\mu},\,B_{a\mu} \,,
\end{eqnarray}
while the allowed fluxes for the NS-NS $B$-field and R-R one-form and 
three-form potentials are $H_{abc}$, $F_{ab}$ and $G_{4abc}$.

The active gauge group ${\Scr G}_g$ is generated by the gauge algebra
$\mathbb{G}_g =\{ X_4 , X, X^a \}$ with connection
\begin{equation}
\varOmega_g = {\Scr G}_\mu^4 X_4 + C_\mu X + B_{a\mu} X^a \,,
\end{equation}
is now purely abelian, even when fluxes are turned on.

On the other hand, the generators of the group ${\Scr G}'_g$ are not
linearly independent, and have the following expressions
\begin{eqnarray}
X^{\prime}_4 &=& G_{4abc}\,
T^{abc}\,,
\nonumber \\
X' &=& H_{abc}\,T^{abc}\,,
\nonumber \\
X^{\prime\,a} &=& F_{bc}\, T^{abc}\,,
\end{eqnarray}
in terms of the generators of the solvable algebra
\begin{eqnarray}
N_4 &=& \{B_{4a}\,T^{ a}+C^{a}\,T_{a}+C^{ab}\,
T_{ab}\}\,,
\nonumber\\
\left[T_{ab},\,T^{c}\right] &=& T_{[a}\delta^c_{b]} \,,
\end{eqnarray}
parametrised by the (non-metric) axionic scalars.

Under the action of ${\Scr G}'_g$ these scalars transform as\footnote{We have 
here defined the scalar $\tilde C ^a = C^a - C^{ab} B_{4a}$, as suggested
by a direct supergravity analysis.}
\begin{eqnarray}
\delta C_{abc} &=& \xi_{[a}\, F_{bc]}+\xi\, H_{abc}+\xi^4\,G_{4abc} \,,
\nonumber\\
\delta B_{4a} &=& 0\,,
\nonumber \\
\delta \tilde{C}^a &=& 0 \,,
\end{eqnarray}
with the only non-trivial covariant derivative given by
\begin{equation}
D_\mu C_{abc} = \partial_\mu C_{abc} - B_{i[a}\, F_{bc]} - C_\mu\,
H_{abc} - {\Scr G}^4_\mu\, G_{4abc} \,.
\end{equation}

\section{Conclusions}

In this paper we have studied the algebraic structure of four-dimensional
$T_6 /\mathbb{Z}_2$ orientifolds, extending the analysis in
\cite{Angelantonj:2003rq}. In the IIA case the active gauge algebras have
dimensions twelve, nine and seven for $p=8,6$ and 4, and their consistency 
implies the condition $F_2 \wedge H_3 =0$ (for $p\not=4$). While in the $p=8$
case it is trivially satisfied, for $p=6$ it implies a constraint on the 
fluxes, in analogy with the $p=7$ case in type IIB \cite{Angelantonj:2003rq}.

Aside from the $p=4$ orientifold, the active gauge algebras are typically
non-abelian when fluxes are turned on, and, for $p=8$ and 5, they are
central extensions of the solvable algebras $N_p$ generated by the 
Peccei-Quinn symmetries of the (non-metric) axionic scalars.

Furthermore, an interesting structure emerges as far as the graviton 
gauge fields ${\Scr G}_\mu^i$ are concerned. Their generators $X_i$
do not commute ($p\not= 3,4,9$) when $H$-fluxes are turned on,
\begin{eqnarray}
p=5 & & [X_i , X_j ] = \epsilon_{ij} H_a X^a\,,
\nonumber \\
p=6 & & [X_i , X_j ] = H_{ij a} X^a + G_{ijab}X^{ab}\,,
\nonumber \\ 
p=7 & & [X_i , X_j ] = H_{ij}^{\varLambda} X_\varLambda \,,
\nonumber \\
p=8 & & [X_i , X_j ] = H_{ij9} X^9 + G_{ijk9} X^{k9} \,,
\end{eqnarray}
independently of our choices of the R-R fluxes.
Since the $X_i$ are four-dimensional remnants of torus translations, 
this signals the non-commutative nature of the torus 
\cite{Connes:1997cr,Seiberg:1999vs} in the presence of $H$ fluxes for
the NS-NS $B$-field.

\vskip 24pt

\noindent
{\bf Acknowledgements.} 
C. A. would like to acknowledge the Physics Department of the
University of Crete for hospitality while completing this paper.
M. T. is grateful to H. Samtleben for useful discussions.
The work of M.T. is supported by an European  Community Marie Curie
Fellowship under contract HPRN-CT-2001-01276, and in part by
EEC under TMR contract ERBFMRX-CT96-0090.
The work of S.F. has been supported in part by European Community's
Human Potential Program under contract HPRN-CT-2000-00131 Quantum
Space-Time, in association with INFN Frascati National Laboratories and 
by D.O.E. grant DE-FG03-91ER40662, Task C.

\end{document}